\documentclass[twocolumn,showpacs,preprintnumbers,amsmath,amssymb]{revtex4}

\usepackage{graphicx}
\usepackage{dcolumn}
\usepackage{bm}

\begin{document}


\title{Dilemma game in a traffic model with the crossing\footnote{W. Zhang and W. Zhang contributed equally to this work.}}

\author{Wei Zhang}
 \email{tzwphys@jnu.edu.cn; wzhang2007065@gmail.com}
\author{Wei Zhang}%
 \email{twzhang@jnu.edu.cn}
\affiliation{%
Department of Physics, Jinan University, Guangzhou 510632, China\\
}%
\date{\today}

\begin{abstract}
In this paper, we investigate the non-signalized intersection issue
considering traffic flow and energy dissipation in terms of game
theory based on the Nagel-Schreckenberg (NaSch) model. There are two
types of driver agents at the intersection when vehicles on the two
streets are approaching to it simultaneously: C agents (cooperative
strategy) pulling up to avoid collision and D agents (defective
strategy) crossing the intersection audaciously. Phase diagram of
the system, which describes free-flow phase, segregation phase,
jammed phase and maximum current curve representing the social
maximum payoff, is presented. Dilemma game is observed at the
phase-segregated state except for the case of $v_{\max }=1.$
\end{abstract}

\maketitle

\section{\label{sec:level1}INTRODUCTION}

Recently, traffic problems have attracted much attention of a
community of physicists because of the observed nonequilibrium phase
transitions and various nonlinear dynamical phenomena. In order to
investigate the dynamical behavior of the traffic flow, a number of
traffic models such as fluid dynamical models, gas-kinetic models,
car-following models and cellular automata (CA)models\cite{1,2,3,4}
have been proposed. These dynamical approaches represented complex
physical phenomena of traffic flow among which are hysteresis,
synchronization, wide moving jams, and phase transitions, etc. Among
these models, the advantages of CA approaches, which have been
extensively applied and investigated, show the flexibility to adapt
complicated features observed in real vehicular traffic\cite{1,4,5}.
The Nagel-Schreckenberg (NaSch) model is a basic CA models
describing one-lane traffic flow\cite{6}. Based on the NaSch model,
many CA models have been extended to investigate the properties of
the system with realistic traffic factors such as highway junctions,
crossing, tollbooths and speed limit zone\cite{1,4,7,8,9,10}.

Previously, scholars pay more attention to traffic flow while
investigating vehicular traffic issues. Most recently, the problems
of energy dissipation in traffic system have been investigated
widely\cite{11,12,13,14,15,16,17} for environmental pollution and
energy dissipation caused by vehicular traffic have become more and
more significant in modern society. Intersections are fundamental
units of complex city traffic networks. Optimization of traffic flow
and energy consumption at a isolated intersection is a substantial
ingredient for the task of global optimization of city networks.
During the past ten years, physicists have paid notable attention to
controlling traffic flow at intersections\cite{8,18,19,20,21,22}.
However, to our knowledge, none of these previous studies about
intersections issue present energy dissipation information, which
should be further investigated.

Signal control works only for major intersections but in most cases,
signal system is not installed due to cost. Drivers at the
intersection without signal system can only communicate each other
by eye contact and make a decision based on own judgment. Most of
the previous studies focus only on the kinetics of the self-driven
multi particle system and ignore the effect of drivers' decisions on
the entire system. In this paper, considering traffic flow and
energy dissipation, we add a game theory framework\cite{23,24,25} as
a rational decision process to the traffic model with a
non-signalized intersection, and demonstrate that the intersection
has a dilemma structure. In addition, the phase diagram which shows
the social maximum payoff is presented.

The paper is organized as follows. Section II is devoted to the
description of the problem. In section III, the results of the
numerical experiment are given and discussed. Finally, the
conclusions are given in section IV.

\section{DESCRIPTION OF THE PROBLEM}

In this section, we present a CA model with two perpendicular one
dimensional closed chains. The chains represent urban streets
accommodating unidirectional vehicular traffic flow. The direction
of traffic flow in the first chain is from south to north and from
east to west in the second chain, as shown in figure 1. Each street
consists of $L$ cells of equal size numbered by $i=1,$ $2,$ $\cdots
,$ $L$ and the time is discrete. The two chains intersect each other
at the sites $i_1=i_2=L/2$ on the first and second chains
respectively. Each site can be either empty or occupied by a vehicle
with the speed $v=0,$ $1,$ $2,\cdots $ $,$ $v_{\max }$, where
$v_{\max }$ is the speed limit. Let $x(i,t)$ and $v(i,t)$ denote the
position and the velocity of the $i$th car at time $t$,
respectively. The number of empty cells in front of the $i$th
vehicle is denoted by $d(i,t)=x(i+1,t)-x(i,t)-1$ . The evolution
dynamics is based on the Nagel-Schreckenberg (NaSch) model. The
updating rules of the NaSch model are as follows:

(1) Acceleration:

$v(i,t+1/3)\rightarrow \min [v(i,t)+1,v_{\max }];$

(2) Slowing down:

$v(i,t+2/3)\rightarrow \min [v(i,t+1/3),d(i,t)];$

(3) Stochastic braking:

$v(i,t+1)\rightarrow \max [v(i,t+2/3)-1,0]$ with the probability
$p;$

(4) Movement: $x(i,t+1)\rightarrow x(i,t)+v(i,t+1).$

The above four steps for all vehicles update in parallel with
periodic boundary.

Vehicles without interactions of vehicles on the perpendicular
streets evolve under the NaSch dynamics. However, how does the
vehicle approaching to the intersection evolve when vehicle on the
other street approaches to the intersection simultaneously? The
approaching driver to the intersection need considering not only the
condition of it's front vehicle but also the situation of the
approaching vehicle on the perpendicular street. Different drivers
perform differently even at the same condition. The decision-making
process of the driver approaching to the intersection is described
by game theory, i.e., we assume that drivers have a strategy that is
either cooperative or defective. Cooperative drivers (C agents) pull
up in the front of the intersection to avoid collision. Defective
drivers (D agents) cross the intersection audaciously. At the same
time, if the two drivers approaching to the intersection on the two
streets are all D agents, i.e., the two drivers adopt defective
strategy simultaneously, traffic accident would occur. Different
from ''the prisoners' dilemma''\cite{26}, one may find that
''non-tit-for-tat'' (I'll cooperate (defect) with you if you defect
(cooperate) with me) is a comparatively effective strategy for
playing the drivers's dilemma. During the simulation, to avoid
collision, we assume that at the same time if the driver approaching
to the intersection on the first street is D agent (C agent), the
driver approaching to the intersection on the second street adopts
cooperative strategy (defective strategy). The probability of the
situation that two approaching drivers to the intersection are D
agents or C agents to occur is very small. In most cases of real
traffic, only one approaching driver to the intersection is D agent
and the other is C agent. Let $P_d$ denotes the probability that the
driver approaching to the intersection on the first street adopt
defective strategy when the other driver on the second street is
also approaching to the intersection at the same time.

\begin{figure}
\includegraphics[height=6cm]{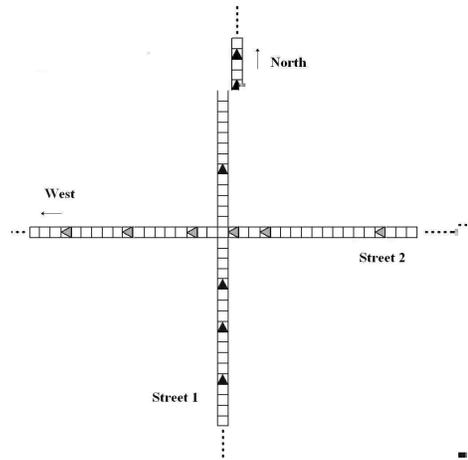}
\caption{\label{fig:epsart} Illustration of the intersection of two
uni-directional streets with periodic boundary. They cross each
other at halfway.}
\end{figure}

The payoff indicates traffic flow $J$ which is the product of the
mean velocity and vehicle density. Except for traffic flow, energy
problem is an important issue in traffic system. The kinetic energy
of the vehicle with the velocity $v$ is $mv^2/2$, where $m$ is the
mass of the vehicle. When braking the kinetic energy reduces. Let
$E_d$ denotes energy dissipation rate per time step per vehicle. For
simple, we neglect rolling and air drag dissipation and other
dissipation such as the energy needed to keep the motor running
while the vehicle is standing in our analysis, i.e., we only
consider the energy lost caused by speed-down. The dissipated energy
of $i$th vehicle from time $t-1$ to $t$ is defined by

\[
e(i,t)=%
{\frac m2\left[ v^2(i,t-1)-v^2(i,t)\right] \quad \text{for }v(i,t)<v(i,t-1) \atopwithdelims\{. 0\qquad \qquad \qquad \qquad \qquad ~~\text{for }v(i,t)\geqslant v(i,t-1).}%
\qquad \left( 1\right)
\]
Thus, the energy dissipation rate

\[
E_d=\frac 1T\frac 1N\sum_{t=t_0+1}^{t_0+T}\sum_{i=1}^Ne(i,t),\qquad
\left( 2\right)
\]
where $N$ is the number of vehicles in the system and $t_0$ is the
relaxation time, taken as $t_0=1.5\times 10^4$. In this model, the
particles are ''self-driven'' and the kinetic energy increases in
the acceleration step. In the stationary state, the value of the
increased energy while accelerating is equivalent to that of the
dissipated energy caused by speed-down, and the kinetic energy is
constant in the system.

In the simulation, the system size $L=500$ and $N_1=N_2$ are
selected where $N_1$ ($N_2$) is the number of vehicles on the first
street (second street) , and the stochastic braking is not
considered, i.e., $p=0$. The numerical results are obtained by
averaging over 20 initial configurations and $5\times 10^3$ time
steps after discarding $1.5\times 10^4$ initial transient states.

\section{NUMERICAL RESULTS}

First of all, we investigate the influences of the drivers' decision
on the social average payoff based on the deterministic NaSch model
with the speed limit $v_{\max }=5$. Figure 2 shows the average
payoff $J$ as a function of the vehicle density $\rho $ with the
probability 0$\leqslant P_d\leqslant 0.5.$ Because of the
equivalence of the two streets, the condition inverts while
0.5$\leqslant P_d\leqslant 1.$ As shown in Fig. 2, there is a
critical density $\rho _{c1}=0.0625$ below which $P_d$ has no
influence on social average payoff and $J$ increases linearly with
the vehicle density. Above the critical density $\rho _{c1,}$ $J$
undergoes a short rapid increase or decrease after which the plateau
arises whose height and length are determined by the probability
$P_d.$ After the plateau, $J$ exhibits linear decrease with the
increase of vehicle density. The intersection of the two streets
makes the crossing point appear as a sidewise dynamical defective
site. The localized defect has global effects whereby the traffic
exhibits macroscopic phase segregation into low-density and
high-density regions. For the first street, the smaller the
probability $P_d$ is, the stronger is the dynamic defect.
Considering the payoff of each street, the larger the probability
that an approaching driver to the intersection adopts defective
strategy, the larger the payoff of the driver and the street on
which the agent drivers are. However, for different vehicle density,
the maximum social mean payoff corresponds to different values of
$P_d.$

Except for $\rho _{c1},$ there are two critical density $\rho
_{c2}=0.167$ below which $J$ is largest in the case of $P_d=0.0$ and
$\rho _{c3}=0.67$ above which $J$ is largest in the case of
$P_d=0.5$. For the whole density region, the maximal social average
payoff $J_{\max }=v_{\max }/2(v_{\max }+1)$ appears at the critical
density $\rho _{c2}.$

\begin{figure}
\includegraphics[height=6cm]{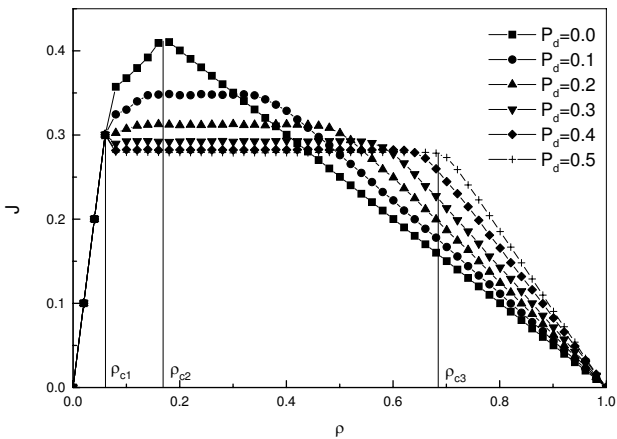}
\caption{\label{fig:epsart} The social average payoff $J$ as a
function of the vehicle density $\rho $ in the case of $v_{\max }=5$
and $p=0$ for various values of the probability $P_d.$ The social
average payoff indicates the mean current of the traffic system with
a single intersection.}
\end{figure}

\begin{figure}
\includegraphics[height=6cm]{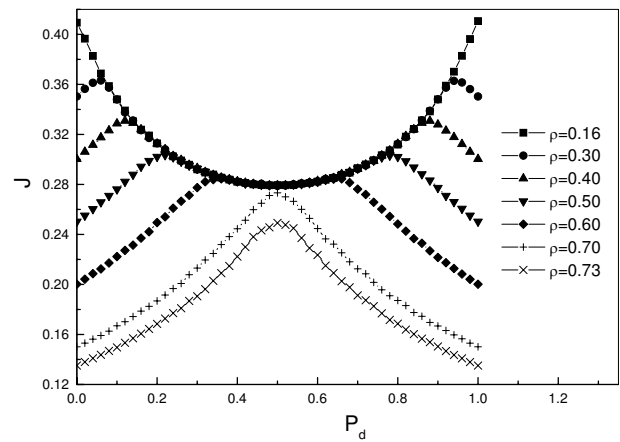}
\caption{\label{fig:epsart} The social average payoff $J$ as a
function of the probability $P_d$ in the case of $v_{\max }=5$ and
$p=0$ for various values of the vehicle density $\rho .$}
\end{figure}

Figure 3 exhibits the relation of $J$ to the probability $P_d$ with
various values of the vehicle density $\rho $ in the case of
$v_{\max }=5$. As expected, the symmetry center of the curve is at
$P_d=0.5$ for the two streets are equivalent in our model. While
$\rho >\rho _{c3}$, with the increase of the probability $P_d,$ $J$
first increases and then decreases after a maximum value is reached.
In the density interval $\rho _{c2}<$ $\rho <\rho _{c3}$, $J$
increases with $P_d$ to the maximum, then it decreases with $P_d$
until $P_d=0.5.$ After the point, $J$ exhibits an increase and
decreases subsequently after the maximum is reached. In the density
interval $\rho _{c1}<$ $\rho <\rho _{c2},$ with the increases of the
probability $P_d,$ $J$ first decreases and increases after a minimum
value is reached. It is noted that when $\rho _{c1}<$ $\rho <\rho
_{c2}$ at the probability $P_d$ interval that the two maximal $J$
appears, curves collapse into one curve which is the social payoff
at $\rho _{c1}$.

Except for fundamental diagram ($J$ versus $\rho $), it is
worthwhile to investigate the energy dissipation diagram of the
system with a intersection. Figure 4 shows the relation of energy
dissipation rate $E_d$ to the vehicle density $\rho $ with various
values of $P_d$ in the case of $v_{\max }=5$. As shown in Fig. 4,
there are three critical density $\rho _{c1}$ below which no energy
dissipation occurs, $\rho _{c2}$ at which there is no energy
dissipation in the case of $P_d=0.0,$ and $\rho _{c3}$ above which
$E_d$ is largest when $P_d=0.5.$ While $\rho \rightarrow \rho
_{c1},$ with the decrease of $P_d,$ energy dissipation rate $E_d$
reduces. And while $P_d=0.0,$ $E_d$ is minimal in the density
interval $\rho _{c1}<$ $\rho <\rho _{c2}$, which is contrary to
traffic flow $J.$ The value of $E_d$ decreases as $P_d$ increases in
the middle density region, but increases in the high density region.

\begin{figure}
\includegraphics[height=6cm]{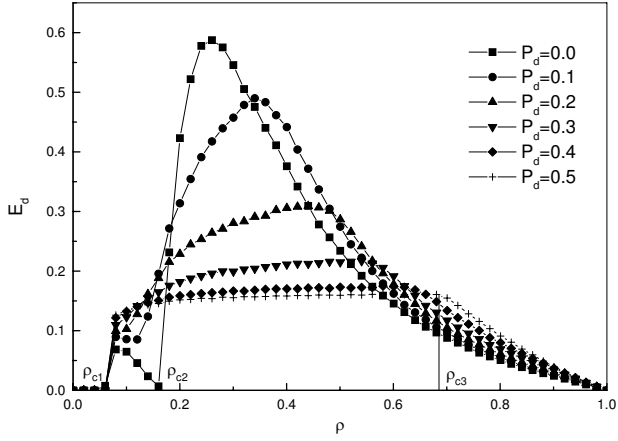}
\caption{\label{fig:epsart} Energy dissipation rate $E_d$ (scaled by
$m$ ) as a function of the vehicle density $\rho $ in the case of
$v_{\max }=5$ and $p=0$ for various values of the probability
$P_d.$}
\end{figure}

From the viewpoint of individual benefit, adopting the higher payoff
strategy is more rational than using the opposite strategy. For
agents on the first street, the larger the $P_d$ is, the more payoff
they obtain. For agents on the second street, the smaller the $P_d$
is, the more payoff they acquire. When the system reaches
equilibrium state, the probability that drivers approaching to the
intersection adopt defective strategy is $0.5.$ However, when
$P_d=0.5$ the average social payoff is not maximum, but minimum in
the density interval $\rho _{c1}<$ $\rho <\rho _{c3}$, which is a
social dilemma.

From Fig. 2 and 4, one should noted that in the density interval
$\rho _{c1}< $ $\rho <\rho _{c2}$, if $P_d=0.0$ i.e., drivers on the
first street (second street) are all C agents (D agents), the payoff
of the whole system is maximal and energy dissipation is minimal.
The best situation having high social efficiency is that drivers on
one of the two streets always pull up and let drivers on the other
street cross, while the interactions of vehicles on the two streets
emerges. However, C agents pulling up for a long time has a robust
incentive to adopt defective strategy for D agents can obtain higher
payoff than C agents. Thus, the probability $P_d$ always increases,
finally reaching absorbed equilibrium $P_d=0.5,$ where the
probability that D agents appear on the two streets is the same.

When $\rho >\rho _{c3}$, the internal equilibrium point $P_d=0.5$ is
consistent with the social maximum payoff where no social dilemma
occurs. This implies that in the jam state D agents appearing on the
two streets with the same probability can improve flow efficiency of
the system, rather than only drivers on the second street adopt
defective strategy.

\begin{figure}
\includegraphics[height=6cm]{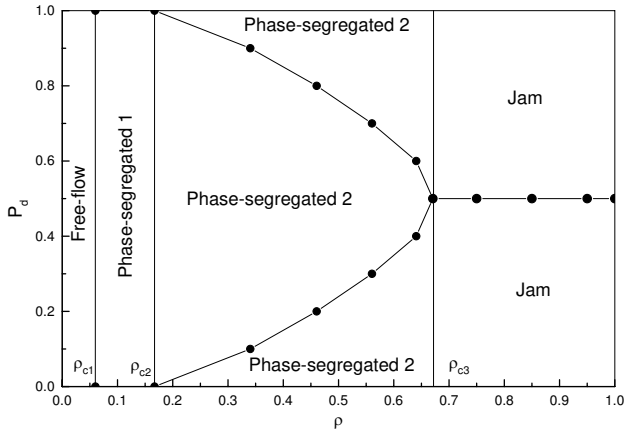}
\caption{\label{fig:epsart} Illustration of the $\rho $-$P_d$ phase
diagram in the case of $v_{\max }=5$ and $p=0$. The closed circles
in the diagram represent the maximum payoff for different vehicle
density.}
\end{figure}

\begin{figure}
\includegraphics[height=6cm]{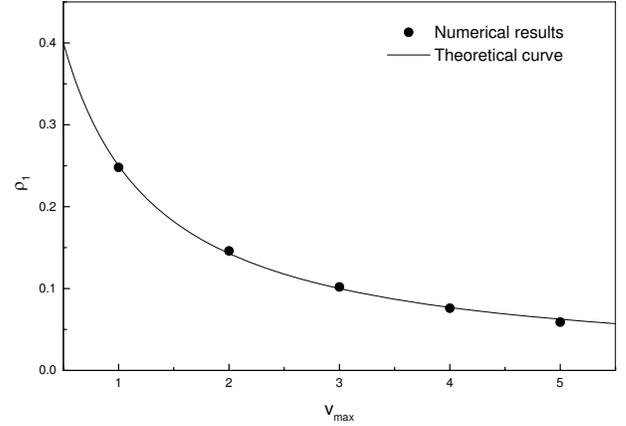}
\caption{\label{fig:epsart} The critical density $\rho _{c1}$ as a
function of the speed limit $v_{\max }$ in the case of $p=0$. Symbol
data are obtained from computer simulations, and solid line
corresponds to analytic results of the formula (3).}
\end{figure}

Figure 5 shows the $\rho $-$P_d$ phase diagram with the speed limit
$v_{\max }=5$. There are four traffic phases: free-flow,
phase-segregated 1, phase-segregated 2 and jammed phase, which are
separated by the critical density $\rho _{c1}$, $\rho _{c2}$ and
$\rho _{c3}$, as shown in Fig. 5. The solid circle symbols in Fig.5
represent the maximum social payoff. In the free-flow phase in which
vehicles can move freely, there are no interactions of vehicles in
the system and no energy dissipation to occur. In the
phase-segregated region, the macroscopic traffic phase segregates
into high-density and low-density region. In the phase-segregated 1
state, the maximal current consists with the minimal energy
dissipation rate. The maximum social payoff appears at $P_d=0.0$ or
$P_d=1.0.$ In the phase-segregated 2 state, however, the maximal
current consists with the maximal energy dissipation rate and $E_d$
increases with traffic flow $J.$ The probability $P_d$ consistent
with the maximal current increases exponentially with the increase
of vehicle density while $P_d<0.5.$ However, when $0.5<P_d<1.0,$
$P_d$ consistent with the maximal current decreases with the
increase of $\rho .$ In the jammed phase, the probability $P_d$
consistent with the maximal current is independent of $\rho $ and
equals 0.5.

Next, we quantitatively analyze the critical density $\rho _{c1}$,
$\rho _{c2}$ and $\rho _{c3}$ for different speed limit $v_{\max }$.
While the mean distance-headway is greater than 3$v_{\max }+1$,
there are no interactions of vehicles in the system and vehicles on
the perpendicular streets can move freely. Consequently, the
critical density $\rho _{c1}$ below which vehicles move freely and
no energy dissipation occurs, can be written as

\[
\rho _{c1}=\frac 1{3v_{\max }+1}.\qquad \qquad (3)
\]

Figure 6 exhibits the relation of the critical density $\rho
_{c1}$to the speed limit $v_{\max }.$ Formula (3) gives an agreement
with numerical results in Fig.6. At the critical density $\rho
_{c2},$ vehicles on the second street (first street) can move freely
in the case of $P_d=0.0$ ($P_d=1.0$). Thus, the critical density
$\rho _{c2}$ is given as

\[
\rho _{c2}=\frac 1{v_{\max }+1}.\qquad \qquad (4)
\]

Above the critical density $\rho _{c3},$ high-density region expands
into the whole system. The critical point $\rho _{c3}$ is not
determined by the speed limit $v_{\max }$ and in the case of
$P_d=0.5,$ $\rho _{c3}$ can be written as

\[
\rho _{c3}=\frac{1/P_d}{1+1/P_d}=\frac 32.\qquad \qquad (5)
\]

However, for the case of $v_{\max }=1$, traffic flow and energy
dissipation exhibit different features. As shown in Fig. 7, there is
only one plateau whose height value is equal to 0.25 for different
values of $P_d.$ The length of the plateau increases with the
increase of $P_d,$ for 0.0$<P_d\leqslant 0.5.$ In the case of
$P_d=0.0,$ there is no plateau and the maximal $J,$ whose value
equals to 0.25, appears at the critical density point $\rho _{c2},$
above which $J$ exhibits linear decrease with the increase of
vehicle density. After the critical density $\rho _{c1},$ the social
average payoff $J$ is maximum in the case of $P_d=0.5$ and is
minimum in the case of $P_d=0.0.$ Consequently, there is no social
dilemma in the case of $v_{\max }=1$.

In the plateau region, for the case of $v_{\max }=1$, the
probability that a approaching vehicle on the first street crosses
the intersection per time step is $P_d/2$ and is (1-$P_d)/2$ on the
second street. Thus, the social average payoff in the plateau region
is 0.25 and does not depend on $P_d.$

\begin{figure}
\includegraphics[height=6cm]{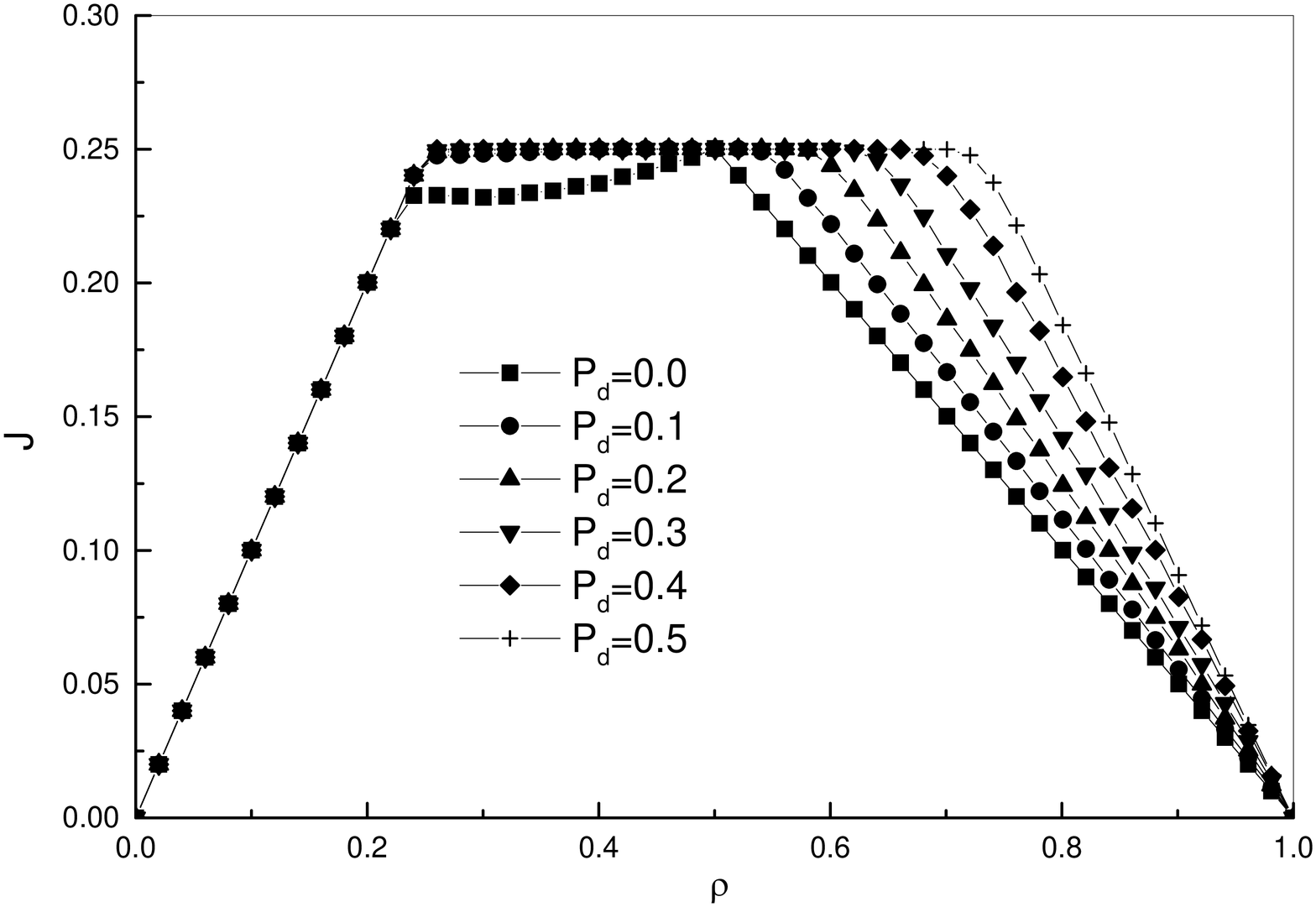}
\caption{\label{fig:epsart} The social average payoff $J$ as a
function of the vehicle density $\rho $ in the case of $v_{\max }=1$
and $p=0$ for various values of the probability $P_d.$}
\end{figure}

\begin{figure}
\includegraphics[height=6cm]{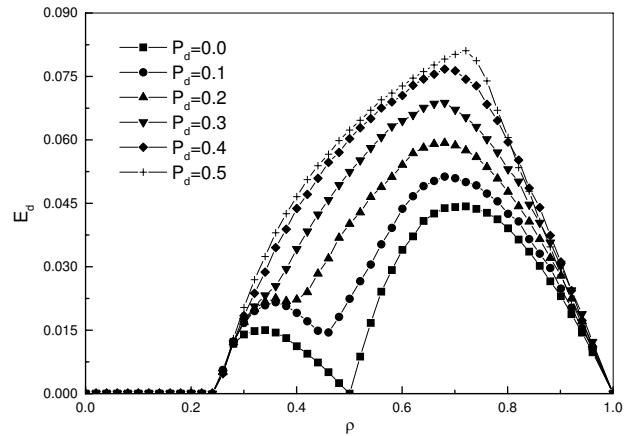}
\caption{\label{fig:epsart} Energy dissipation rate $E_d$ (scaled by
$m$ ) as a function of the vehicle density $\rho $ in the case of
$v_{\max }=1$ and $p=0$ for various values of the probability
$P_d.$}
\end{figure}

Figure 8 shows energy dissipation rate $E_d$ as a function of the
vehicle density $\rho $ with various values of the probability $P_d$
in the case of $v_{\max }=1$. As shown in Fig.8, there are two
critical density $\rho _{c1}$ below which no energy dissipation
occurs, and $\rho _{c2}$ at which no go-and-stop vehicles appear
when $P_d=0.0.$ The energy dissipation rate $E_d$ increases with the
increase of $P_d$ for 0.0$\leqslant P_d\leqslant 0.5,$ which is
different from that for $v_{\max }>1$. Considering the $\rho $-$P_d$
phase diagram, there is no differences between phase-segregated 1
and phase-segregated 2, and $P_d$ consistent with the maximal social
payoff is always equal to 0.5 and independent of $\rho $ in
phase-segregated and jammed states (not shown). Therefore these
results indicate that different correlations of spacetime exist
between the case of $v_{\max }=1$ and $v_{\max }>1.$

\section{SUMMARY}

In this paper, we investigated the social dilemma structure in a
traffic model with a non-signalized intersection based on the NaSch
model. The model contains a game theory framework to deal with
drivers' decision-making processes. We studied the effects of the
drivers' decision on traffic flow and energy dissipation at
different traffic phases.

Numerical results indicate that in the case of $v_{\max }>1$ the
social dilemma appears at the phase-segregated states and no dilemma
exists at other traffic phases. At the phase-segregated states,
selfish drivers crossing the intersection can obtain a higher payoff
than altruistic drivers pulling up in the front of the intersection,
but they cause a remarkable decrease in social efficiency when they
emerge alternately on the two streets. In contrast to the
phase-segregated states, in the jammed phase, the social efficiency
is maximal at the absorbed equilibrium $P_d=0.5.$ Different from
that in the case of $v_{\max }>1,$ in the case of $v_{\max }=1,$
there is no dilemma to occur no matter in the phase-segregated and
jammed states.

In addition, the three critical density $\rho _{c1},$ $\rho _{c2}$
and $\rho _{c3}$ were analyzed quantitatively and theoretical
analyses give an excellent agreement with numerical results.
However, explicit expressions about the maximum social payoff curve
in the $\rho $-$P_d$ phase diagram do not be obtained because of
effects of long length of time space correlations, and deserve
further investigate.

\end{document}